\documentclass[a4paper,11pt]{article}
\usepackage{jheppub}
\arxivnumber{2602.19363}
\usepackage[T1]{fontenc}
\usepackage[utf8]{inputenc}
\usepackage{lmodern}           
\usepackage{microtype}         
\usepackage{amsmath,amssymb,bm,mathtools}
\usepackage{siunitx}
\usepackage{graphicx}
\usepackage{booktabs,multirow,tabularx}
\usepackage[svgnames,x11names,dvipsnames,table]{xcolor}
\usepackage[nameinlink,noabbrev]{cleveref}
\usepackage{float}

\usepackage{tikz}
\usetikzlibrary{arrows.meta,positioning,shapes.geometric,fit,calc,backgrounds,decorations.pathreplacing}

\definecolor{CurveAmber}{HTML}{B45309}
\definecolor{CurveBlue}{HTML}{1E88E5}
\definecolor{CurveRed}{HTML}{E53935}
\definecolor{CurveGreen}{HTML}{2E7D32}
\definecolor{CurveOrange}{HTML}{F39C12}

\graphicspath{{./}{/mnt/data/}}

\title{Certified Uncertainty for Surrogate Models of Neutron Star Equations of State via Mondrian Conformal Prediction}

\author[a]{Marlon M. S. Mendes}
\author[a]{Roberta Duarte Pereira}
\author[a]{Mariana Dutra}
\author[a]{C\'esar H. Lenzi}

\affiliation[a]{Departamento de F\'isica e Laborat\'orio de Computa\c c\~ao Cient\'ifica Avan\c cada e Modelamento (Lab-CCAM), Instituto Tecnol\'ogico de Aeron\'autica, DCTA, 12228-900, S\~ao Jos\'e dos Campos, SP, Brazil}

\emailAdd{marlon.mendes.102043@ga.ita.br}

\abstract{
We present a multitask surrogate for neutron-star equations of state (EoSs) that delivers
\emph{distribution-free}, certified uncertainty via split conformal prediction (CP) and its Mondrian
variant. The surrogate ingests a six-parameter piecewise--polytropic representation
$(\log_{10}p_1,\Gamma_1,\Gamma_2,\Gamma_3,\rho_1,\rho_2)$---with fixed transition densities $\rho_1$ and
$\rho_2$---and jointly performs (i) validity classification under physical/observational constraints and
(ii) regression of $M_{\max}$, $R(M_{\max})$, $R_{1.4}$, and $\Lambda_{1.4}$. Trained on a balanced set of
$40{,}000$ EoSs, the model attains near-perfect discrimination (AUC $\approx 0.997$) and sub-percent
relative errors for masses and radii, with few-percent error for tidal deformability. Across
$\alpha\in[0.05,0.25]$, empirical coverages closely track $1-\alpha$ for both Standard and Mondrian CP;
in conservative regimes, Mondrian yields narrower average physical widths at comparable coverage. To our knowledge, this is the first application of class-conditioned (Mondrian) conformal
calibration to neutron-star EoS surrogates, enabling efficient, reproducible, and uncertainty-aware
inference; the framework is readily extensible to functional targets (e.g., full $R(M)$ curves).
}


\begin{document}
\maketitle


\section{Introduction} \label{sec:intro}

The neutron star (NS) equation of state (EoS) encodes the fundamental relation between pressure and energy density in ultra-dense matter, directly governing macroscopic observables such as its mass and radius, in addition to the tidal deformability \citep{lattimer2012,ozel2016}. Constraining the EoS from multi-messenger observations — including electromagnetic signals and gravitational waves (GW) — requires repeated solutions of the Tolman–Oppenheimer–Volkoff (TOV) equations \citep{tolman1939,oppenheimer1939}, in conjunction with the tidal deformability equations, across large sets of candidate EoSs. This process can be computationally expensive, particularly when coupled with Bayesian inference pipelines or large-scale parameter sweeps. These aspects provide the motivation for the development of surrogate models that emulate TOV integration while preserving physical consistency \citep{fujimoto2020,landry2020b}. 

A central challenge for such surrogates is providing \emph{certified} uncertainty quantification (UQ) that remains valid under minimal assumptions. In current NS astrophysics literature, UQ has been pursued mainly through Bayesian neural networks (BNNs) \citep{grigorian2018,landry2020} and Gaussian processes \citep{stein2021}, as well as posterior sampling frameworks informed by nuclear theory and GW observations \citep{annala2018,essick2020}. While these methods are powerful, they typically depend on restrictive distributional priors, scale poorly in high-dimensional parameter spaces, and do not provide finite-sample, distribution-free guarantees.

Conformal Prediction (CP) \citep{vovk2005,angelopoulos2021} addresses these limitations by wrapping any black-box predictor with statistically rigorous coverage guarantees under the mild assumption of data exchangeability. CP is model-agnostic, computationally lightweight, and naturally applicable to both regression and classification tasks. Variants such as Normalized CP adapt interval sizes to local difficulty estimates, while Mondrian CP \citep{bostrom2020,bostrom2021} stratifies calibration by discrete categories, yielding tighter prediction sets without sacrificing coverage. In addition, the regression-as-classification paradigm \citep{romano2020} extends CP’s flexibility by discretizing target spaces into ordered bins, enabling classification-based CP methods to capture heteroscedastic and multimodal uncertainties.

Recent advances position CP as a general-purpose, distribution-free calibration layer for scientific machine learning. For PDE problems, \emph{conformalized physics-informed neural networks} (C-PINNs) have been introduced \citep{Podina2024_CPINN}.
For operator learning, conformalized operator learners—e.g., \emph{Conformalized-DeepONet}~\citep{Moya2024_CDeepONet} and risk-controlling quantile neural operators \citep{Ma2024_RCQNO},
built upon the original DeepONet architecture \citep{Lu2021_DeepONet}—demonstrate valid and
interpretable uncertainty quantification in high-dimensional physical systems. Yet, despite these
advances, CP has not been applied to neutron-star EoS surrogates, nor to multitask settings that jointly address the regression of continuous observables (e.g., some observables of NS like mass, radii, or tidal deformabilities) and
the classification of discrete EoS attributes (e.g., validity, phase-transition signatures).

This gap is particularly significant because CP’s finite-sample, model-agnostic guaranties can
be conditioned on physically meaningful categories, enabling uncertainty estimates to be
expressed directly in physical units and focused on scientifically critical regimes. In
principle, Mondrian stratification based on stellar mass or compactness could yield narrower
and more informative intervals in both low- and high-mass regions, while adaptive allocation
schemes for the significance level $\alpha$ \citep{fisch2021} could prioritize coverage where
observational constraints are most stringent—such as the NICER credible regions for
PSR~J0030$+$0451 and PSR~J0740$+$6620
\citep{Miller2019_ApJL_J0030,Riley2019_ApJL_J0030,Miller2021_ApJL_J0740,Riley2021_ApJL_J0740}
or the LIGO–Virgo bounds on $\Lambda_{1.4}$
\citep{Abbott2018_PRL_GW170817,Abbott2020_ApJL_GW190425}. Integrating these ideas within a
multitask EoS surrogate offers a principled route to valid coverage, improved efficiency, and
physically interpretable uncertainty bands, thereby strengthening the reliability of EoS
inference in the era of precision neutron-star astrophysics.

For clarity, we briefly outline the structure of the paper. Section~\ref{sec:data} describes
the construction of the EoS dataset, the physical filters applied, and the associated
observational constraints from NICER and gravitational-wave measurements. Section~\ref{sec:model}
presents the surrogate architecture, preprocessing steps, and the multitask formulation used
to predict both scalar observables and EoS validity. Section~\ref{sec:cp} introduces Standard
and Mondrian conformal prediction, detailing the calibration procedure and the resulting
coverage guarantees. Section~\ref{sec:results} evaluates the surrogate through accuracy,
coverage, interval width, and external validation under distributional shift. Section~\ref{sec:benchmarks}
contextualizes the method relative to Bayesian and neural surrogates in the literature, with
emphasis on differences in uncertainty quantification. Finally, Section~\ref{sec:disc_conc}
summarizes the main contributions, discusses limitations, and outlines future directions
including functional surrogates, physics-informed decoders, and conformal calibration in
function space.

\section{Methods}\label{sec:methods}

\subsection{Dataset}\label{sec:data}

\begin{figure}[H]
\centering
\includegraphics[width=0.8\columnwidth]{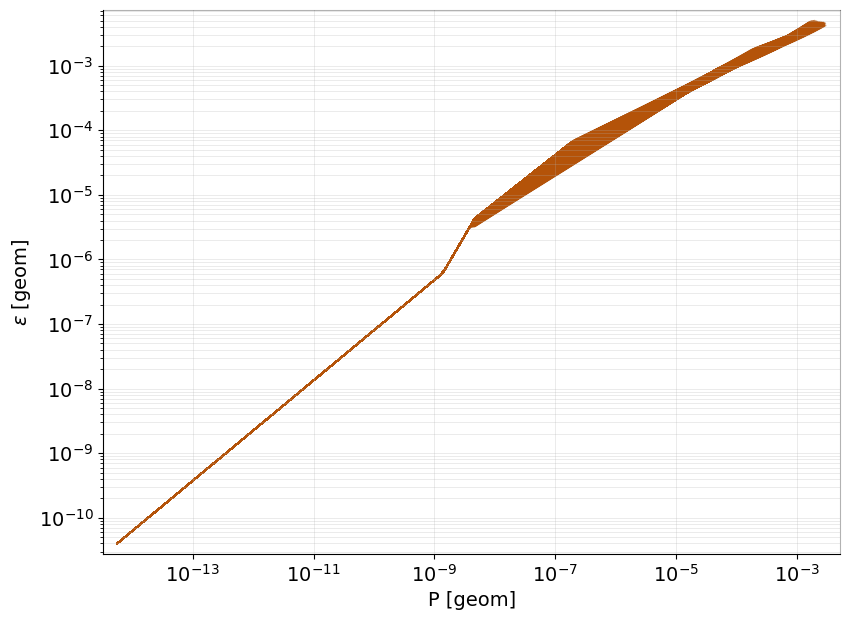}
\caption{\footnotesize
Energy density versus pressure $\varepsilon(P)$ (log--log) for the valid EoS subset.
Each amber trace corresponds to one EoS realization that satisfies our physical
filters (stability, causality, and a lower bound on the maximum mass), illustrating
the diversity of stiffness/softness across the ensemble.
In the axes, $P[\mathrm{geom}]$ and $\varepsilon[\mathrm{geom}]$ denote pressure and
energy density expressed in geometric units, where $G = c = 1$; in this system
they scale as inverse length squared and relate to cgs units via
$P_{\mathrm{geom}} = (G/c^{4})\,P_{\mathrm{cgs}}$ and
$\varepsilon_{\mathrm{geom}} = (G/c^{4})\,\varepsilon_{\mathrm{cgs}}$.
}
\label{fig:eos_epsP}
\end{figure}

\begin{figure}[H]
\centering
\includegraphics[width=0.8\columnwidth]{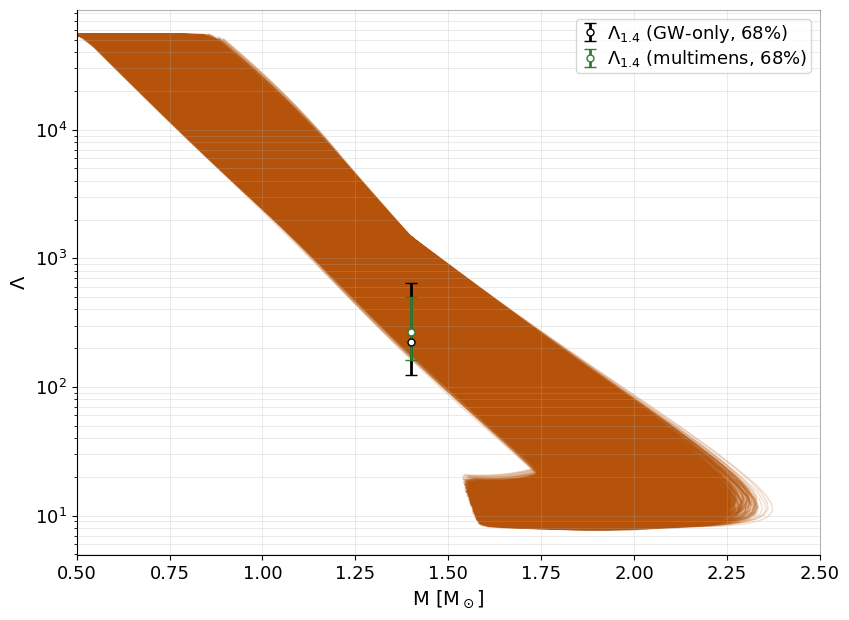}
\caption{\footnotesize
Tidal deformability curves $\Lambda(M)$ for $0.5\!\le\!M/M_\odot\!\le\!2.5$.
The amber background shows the valid EoS ensemble, while the vertical band at
$M=1.4\,M_\odot$ highlights two EOS-agnostic constraints:
a GW-only estimate $\Lambda_{1.4}=222.9^{+420.3}_{-98.9}$ and a multimessenger estimate
$\Lambda_{1.4}=265.2^{+237.9}_{-104.4}$.
}
\label{fig:eos_lambda}
\end{figure}

\begin{figure}[H]
\centering
\includegraphics[width=0.8\columnwidth]{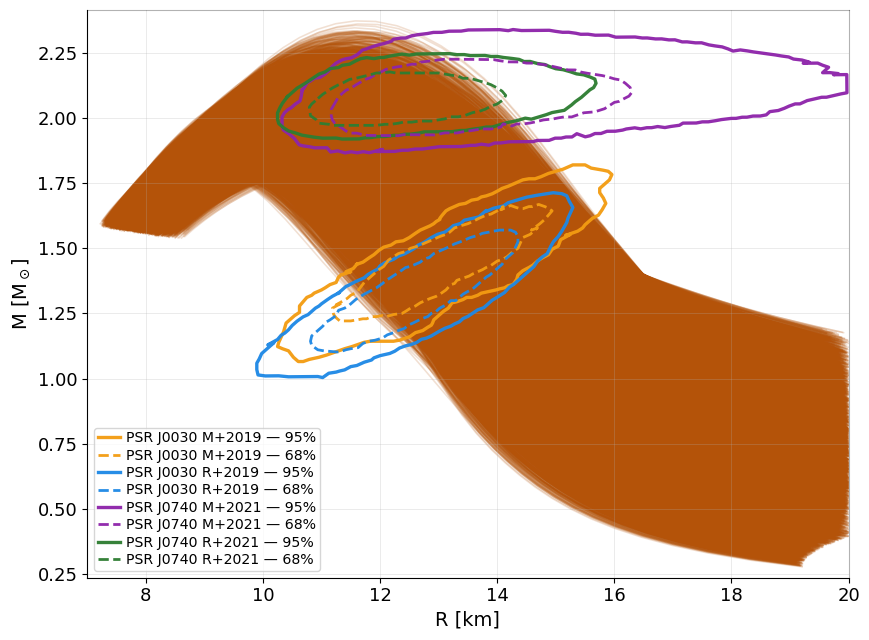}
\caption{\footnotesize
Mass–radius relations $M$--$R$ for the valid EoS subset (amber background),
overlaid with NICER constraints for PSR~J0030+0451 and PSR~J0740+6620.
Outer contour boundaries are solid; inner credible levels are dashed.
}
\label{fig:eos_mr}
\end{figure}

To illustrate the structure and variability of our equation-of-state ensemble,
Figs.~\ref{fig:eos_epsP}--\ref{fig:eos_mr} show three complementary projections of
the valid subset. Figure~\ref{fig:eos_epsP} displays the pressure–energy-density
relation $\varepsilon(P)$, Fig.~\ref{fig:eos_lambda} shows the tidal deformability
curves $\Lambda(M)$, and Fig.~\ref{fig:eos_mr} presents the corresponding mass–radius
relations $M$--$R$, each panel including the relevant observational constraints.

The observational layers overlaid on our curated EoS subset are shown in
Figs.~\ref{fig:eos_epsP}--\ref{fig:eos_mr}.  
In the mass--radius plane (Fig.~\ref{fig:eos_mr}), we include the two independent
NICER inferences for PSR~J0030+0451
\citep{Miller2019_ApJL_J0030,Riley2019_ApJL_J0030} and PSR~J0740+6620
\citep{Miller2021_ApJL_J0740,Riley2021_ApJL_J0740}, plotting their outer credible
contours as solid lines and inner credible levels as dashed.  
In the tidal–deformability panel (Fig.~\ref{fig:eos_lambda}), restricted to
$0.5 \le M/M_\odot \le 2.5$, we mark two EOS-agnostic point estimates at
$M=1.4\,M_\odot$ with asymmetric $68\%$ credible intervals: a GW-only determination,
$\Lambda_{1.4}=222.9^{+420.3}_{-98.9}$, and a multimessenger determination,
$\Lambda_{1.4}=265.2^{+237.9}_{-104.4}$ \citep{Huang2025_EOSagnostic_Lambda14}.
These values are consistent with the constraint envelope derived from GW170817 and
GW190425 \citep{Abbott2018_PRL_GW170817,Abbott2020_ApJL_GW190425}, reproduced in
Fig.~\ref{fig:eos_lambda} for reference.
Across panels, the amber background traces visualize the diversity of the \emph{valid} EoS set used throughout this work.

\begin{equation}
  \varepsilon(\rho) = (1+\alpha_i)\rho + \frac{K_i}{\Gamma_i-1}\rho^{\Gamma_i},
  \label{eq:energy_density}
\end{equation}
where the constants $\alpha_i$ are determined recursively to enforce continuity of $\varepsilon(\rho)$ across each segment boundary.

\paragraph{Notation.}
The coefficients $\alpha_i$ in Eq.~\eqref{eq:energy_density} represent the \emph{piecewise energy–density continuity parameters} that guarantee the thermodynamic smoothness of the EoS at the transition densities. Each $\alpha_i$ is obtained recursively from the continuity condition
\[
  \varepsilon_i(\rho_{i}) = \varepsilon_{i+1}(\rho_{i}),
\]
ensuring that both pressure and energy density remain continuous across the interfaces between polytropic segments. 
Importantly, this $\alpha_i$ notation is unrelated to the miscoverage level $\alpha$ that will later appear in the conformal–prediction framework. 
We emphasize this distinction to prevent confusion between the EoS continuity constants $\{\alpha_i\}$ and the CP significance parameter.

\paragraph{Sampling ranges and units.}
The uniform sampling intervals adopted for each EoS parameter are listed in Table~\ref{tab:sampling_ranges}. 
Pressures are expressed in cgs units and converted to geometric units via 
$p_1^{\rm geom}=10^{p_{1,\log}}/(\mathrm{Density}\,c^2)$. 
At subnuclear densities, the SLy–PP EoS is attached to ensure crustal consistency \citep{Read2009}.

\begin{table}[h!]
\centering
\caption{Uniform sampling ranges and units for the piecewise–polytropic EoS parameters.}
\label{tab:sampling_ranges}
\begin{tabular}{lcc}
\hline
\textbf{Symbol} & \textbf{Range} & \textbf{Units / Notes} \\
\hline
$p_{1,\log_{10}}$ & $[34.4,\,34.8]$ & cgs (converted to geometric) \\
$\Gamma_1$ & $[1.6,\,2.2]$ & dimensionless \\
$\Gamma_2$ & $[2.0,\,2.8]$ & dimensionless \\
$\Gamma_3$ & $[2.2,\,3.3]$ & dimensionless \\
$\rho_1$ & $10^{14.7}$ & g\,cm$^{-3}$ (fixed) \\
$\rho_2$ & $10^{15.0}$ & g\,cm$^{-3}$ (fixed) \\
\hline
\end{tabular}

\vspace{0.5em}
\begin{minipage}{0.95\linewidth}
\footnotesize\textit{Notes.} 
$p_{1,\log_{10}}$ is the base-10 logarithm of the pressure at the first transition density $\rho_1$. 
The triplet $(\Gamma_1,\Gamma_2,\Gamma_3)$ denotes the adiabatic indices governing the stiffness of each polytropic segment. 
The transition densities $\rho_1$ and $\rho_2$ delimit the boundaries between successive segments and are fixed to match the typical nuclear-to-core regime transition identified by \citet{Read2009}. 
The SLy–PP Equation of State combines the microscopic Skyrme–Lyon (SLy4) model for the low-density crust with a piecewise–polytropic (PP) parameterization for the high-density core, ensuring thermodynamic continuity at $\rho_1$ and providing a realistic description of the outer layers of neutron stars. 
This composite EoS scheme has become a standard choice in neutron-star modeling because it offers a smooth and causal connection between the outer crust and the core polytropic segments, while maintaining stability and compatibility with empirical nuclear constraints \citep{Read2009,Haensel2001_SLy}.

\end{minipage}
\end{table}

\paragraph{Geometric units.}
All quantities are expressed in geometric units ($G=c=1$), which absorb constants into the definitions of mass, length, and density.
In this convention, one solar mass corresponds to a length scale $GM_\odot/c^2 \approx 1.47664\,\mathrm{km}$. Densities are normalized as
$\rho = M_\odot/\mathrm{Length}^3$, while pressures and energies share the same units as density. This formulation eliminates cumbersome
factors of $G$ and $c$ in the Tolman–Oppenheimer–Volkoff equations, improving numerical stability and interpretability
\citep{tolman1939,oppenheimer1939}.

\paragraph{Targets and constraints.}
Each candidate EoS is integrated through the TOV equations to extract the target vector
\begin{equation}
  \mathbf{y} = \big(M_{\max},\,R(M_{\max}),\,R_{1.4},\,\Lambda_{1.4}\big),
\end{equation}
where $M_{\max}$ is the maximum gravitational mass, $R(M_{\max})$ its corresponding radius, $R_{1.4}$ the radius of a $1.4\,M_\odot$ star,
and $\Lambda_{1.4}$ the tidal deformability at $1.4\,M_\odot$. The latter is particularly sensitive because
$\Lambda\propto R^5/M^5$, amplifying small uncertainties in $R_{1.4}$ into the high-density regime. To ensure physical consistency, we enforce:
\begin{itemize}
  \item \textbf{Stability:} $\mathrm{d}P/\mathrm{d}\varepsilon > 0$ across all segments;
  \item \textbf{Causality:} $c_s^2 = \mathrm{d}P/\mathrm{d}\varepsilon \leq c^2$,
        with a small tolerance margin ($\lesssim 10$--$15\%$) following \citet{Read2009},
        to account for interpolation errors in piecewise–polytropic fits;
  \item \textbf{Astrophysical viability:} $M_{\max} \geq 2.0\,M_\odot$, consistent with heavy pulsars (e.g., PSR~J0740+6620; \citealt{Cromartie2020_NatAstro, Fonseca2021_ApJL}).
\end{itemize}

\paragraph{Distribution of $\Lambda_{1.4}$.}
The dataset is deliberately constructed to sample a wide range of $\Lambda_{1.4}$ values, spanning $\sim 100$--$1500$.
This broad spread reflects the high sensitivity of tidal deformability to core stiffness and introduces greater variance than seen in
$M_{\max}$ or radii. As a result, conformal prediction (CP) intervals for $\Lambda_{1.4}$ are systematically wider, capturing the intrinsic
uncertainty of the high-density EoS sector. By including both extremely soft and stiff models, the calibration process remains valid across
the full physical range, ensuring that empirical coverage closely tracks the nominal $1-\alpha$ even in this most challenging observable.

EoSs violating any of the constraints are labeled as invalid. The final dataset contains $40{,}000$ EoSs ($20{,}000$ valid and $20{,}000$
invalid), balanced by design to support both classification and regression. The valid subset is used to compute normalization statistics for
the regression targets, ensuring that no information leaks from calibration or test splits. This construction mirrors the philosophy of
\citet{Read2009}, but extends it with explicit causal filters, a balanced invalid class, and a tailored sampling of $\Lambda_{1.4}$ to
facilitate surrogate training under conformal prediction.

\section{Model architecture}\label{sec:model}

\paragraph{Data representation.}
The surrogate is trained on a consolidated archive of $\sim4\times10^4$ EoSs stored
as NumPy arrays, following conventions similar to
\citet{Landry2019_PRD_GP_EoS,Fujimoto2023_ApJ_NN_Surrogate}.
Each sample contains a 6-dimensional feature vector
$X=\{\log_{10}p_1,\Gamma_1,\Gamma_2,\Gamma_3,\rho_1,\rho_2\}$, consistent with
piecewise-polytropic descriptions \citep{Read2009}. Features are
standardized using training-only statistics and the resulting scaler is frozen for
validation, test, and downstream inference.

\paragraph{Targets and masking.}
Two supervision channels are used:
(i) a binary validity label $y_{\mathrm{class}}$ obtained from TOV integration under
stability, causality, and $M_{\max}$ constraints;  
(ii) a 4-vector regression target
$y_{\mathrm{reg}}=\big(M_{\max},R_{\max},R_{1.4},\Lambda_{1.4}\big)$.
A boolean mask gates the regression loss, excluding samples where any scalar is
undefined (e.g., TOV failure or unphysical profiles), while classification remains
active for all EoSs. Following \citet{Fujimoto2023_ApJ_NN_Surrogate}, $\Lambda_{1.4}$
is modeled in $\log$-space for numerical stability.

\paragraph{Splits and calibration.}
We adopt an $80/20$ train/test split, reserving an internal validation subset for
early stopping and hyperparameter tuning. A disjoint calibration subset is used to
fit both Standard and Mondrian Conformal Prediction (CP). All normalizers are fitted
on the training split only.

\paragraph{Network heads.}
A shared representation branches into:
(i) a sigmoid neuron yielding $p(\mathrm{valid}\mid X)$; and  
(ii) four linear neurons outputting standardized regression predictions
$(\hat M_{\max},\hat R_{\max},\hat R_{1.4},\widehat{\Lambda}_{1.4})$.

\paragraph{Loss function.}
The total objective combines classification and masked regression,
\[
\mathcal{L}
= \lambda_{\mathrm{cls}}\,\mathcal{L}_{\mathrm{BCE}}
+ \sum_{\theta\in\Theta}\lambda_\theta\;
\mathbf{1}_{\{\mathrm{mask}=1\}}\,
\mathcal{L}^{(\theta)}_{\mathrm{MSE}},
\qquad
\Theta=\{M_{\max},R_{\max},R_{1.4},\Lambda_{1.4}\}.
\]
Classification is optimized over all samples, whereas each regression head contributes
only when physically meaningful targets exist. This prevents leakage between validity
prediction and the regression gate while ensuring stable multitask learning.

\noindent\textbf{Activation and parameterization.}
Hidden layers use ReLU, $f(x)=\max(0,x)$, for efficient training and expressive nonlinear mappings between microscopic EoS parameters and macroscopic observables. The six polytropic parameters $\mathbf{x}=\{p_{1,\log_{10}},\Gamma_1,\Gamma_2,\Gamma_3,\rho_1,\rho_2\}$ define the piecewise–polytropic EoS following \citet{Read2009}, where $p_{1,\log_{10}}$ is the logarithmic pressure at $\rho_1$, $(\Gamma_1,\Gamma_2,\Gamma_3)$ control the stiffness of each segment, and $(\rho_1,\rho_2)$ are the transition densities separating polytropic intervals.

\noindent\textbf{Operating thresholds and external validation.}
Unless stated otherwise, the default validity threshold is $0.5$; we also report conservative fixed thresholds (e.g., $0.99$) and data-driven choices (e.g., Youden’s $J$) selected on validation. External OOD evaluations use a distinct dataset built after freezing the model and scaler, ensuring zero leakage.

\noindent\textbf{Uncertainty and training details.}
At inference, Monte Carlo dropout may be enabled to approximate epistemic uncertainty; CP calibration (Standard/Mondrian) remains distribution-free and model-agnostic. We train with Adam, learning-rate scheduling (ReduceLROnPlateau), early stopping, and weight decay. Hyperparameters (layer widths, dropout, and per-head weights) are tuned to balance accuracy, stability, and throughput for large-scale sweeps.

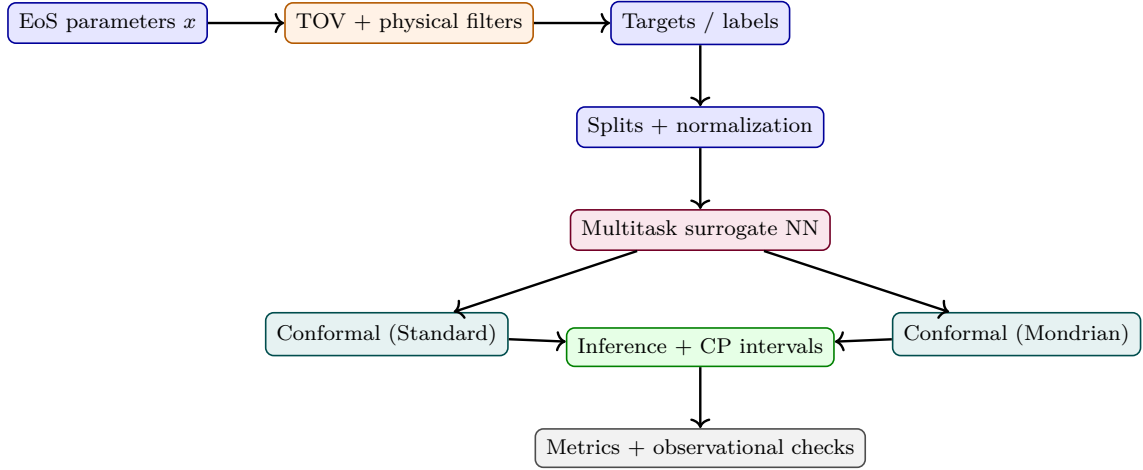
\begin{figure}[H]
\centering
\resizebox{1.1\linewidth}{!}{%
\begin{tikzpicture}[
  node distance=7mm and 10mm,
  font=\scriptsize,
  data/.style  ={rounded corners=3pt, draw=blue!60!black,   fill=blue!10,   semithick, inner sep=4pt, align=left},
  proc/.style  ={rounded corners=3pt, draw=orange!70!black, fill=orange!10, semithick, inner sep=4pt, align=left},
  model/.style ={rounded corners=3pt, draw=purple!60!black, fill=purple!10, semithick, inner sep=4pt, align=left},
  calib/.style ={rounded corners=3pt, draw=teal!60!black,   fill=teal!10,   semithick, inner sep=4pt, align=left},
  outp/.style  ={rounded corners=3pt, draw=green!50!black,  fill=green!10,  semithick, inner sep=4pt, align=left},
  obs/.style   ={rounded corners=3pt, draw=gray!60!black,   fill=gray!10,   semithick, inner sep=4pt, align=left},
  flow/.style  ={->, line width=.9pt}
]

\node[data] (x)   {EoS parameters $x$};
\node[proc, right=10mm of x] (tov) {TOV + physical filters};
\node[data, right=10mm of tov] (y) {Targets / labels};
\draw[flow] (x)--(tov);
\draw[flow] (tov)--(y);

\node[data, below=8mm of y] (split) {Splits + normalization};
\draw[flow] (y)--(split);

\node[model, below=8mm of split] (sur) {Multitask surrogate NN};
\draw[flow] (split)--(sur);

\node[calib, below left=8mm and 8mm of sur]  (std) {Conformal (Standard)};
\node[calib, below right=8mm and 8mm of sur] (mon) {Conformal (Mondrian)};
\draw[flow] (sur)--(std);
\draw[flow] (sur)--(mon);

\node[outp, below=10mm of sur] (inf) {Inference + CP intervals};
\draw[flow] (std)--(inf);
\draw[flow] (mon)--(inf);

\node[obs, below=8mm of inf] (eval) {Metrics + observational checks};
\draw[flow] (inf)--(eval);

\end{tikzpicture}%
}
\caption{Conformal prediction pipeline for a multitask surrogate neural network.}
\label{fig:cp_pipeline_premium}
\end{figure}



\subsection{Conformal prediction}\label{sec:cp}

We quantify predictive uncertainty using \emph{split conformal prediction} (CP)
\citep{vovk2005,angelopoulos2021}, employing a training set for model fitting and a
disjoint calibration set to compute empirical quantiles that guarantee finite-sample
coverage under exchangeability.

\paragraph{Classification.}
For the validity classifier, the nonconformity score is
$s(\mathbf{x},y)=1-\hat{p}_y$, yielding prediction sets
$\mathcal{C}(\mathbf{x}^\ast)=\{y:\,s(\mathbf{x}^\ast,y)\le q_{1-\alpha}\}$.
This provides guaranteed marginal coverage at level $1-\alpha$.

\paragraph{Regression.}
For each scalar target
$\theta\in\{M_{\max},R_{\max},R_{1.4},\Lambda_{1.4}\}$, we use absolute standardized
residuals $s=|y-\hat y|$ on the calibration set. The $(1-\alpha)$ quantile produces a
symmetric interval
$[\hat y(\mathbf{x}^\ast)-q_{1-\alpha},\,\hat y(\mathbf{x}^\ast)+q_{1-\alpha}]$,
later mapped back to physical units. Intervals are computed independently per output.

\paragraph{Mondrian CP.}
To improve conditional validity, we also implement \emph{Mondrian CP}, computing
quantiles within stratified groups (e.g., by EoS stiffness or validity class).
Because error distributions differ across regimes—especially for quantities such as
$\Lambda_{1.4}$, whose variance scales steeply with radius—groupwise calibration
often yields tighter, more informative intervals at fixed nominal coverage.

\paragraph{Astrophysical relevance.}
This framework provides certified, data-adaptive uncertainty quantification for
observables that strongly differentiate between viable EoSs—most notably
$\Lambda_{1.4}$, whose $R^5$ dependence amplifies model errors
\citep{Hinderer2008_ApJ_Tidal,Hinderer2010_PRD_Tidal}. The resulting intervals are
directly comparable to NICER and LIGO/Virgo constraints
\citep{Miller2019_ApJL_J0030,Riley2019_ApJL_J0030,
Miller2021_ApJL_J0740,Riley2021_ApJL_J0740,
Abbott2018_PRL_GW170817,Abbott2020_ApJL_GW190425,Huang2025_EOSagnostic_Lambda14},
allowing robust confrontation of surrogate predictions with observations.

\section{Results}\label{sec:results}
\subsection{Predictive performance}

\begin{figure}[H]
    \centering
    \includegraphics[width=0.8\linewidth]{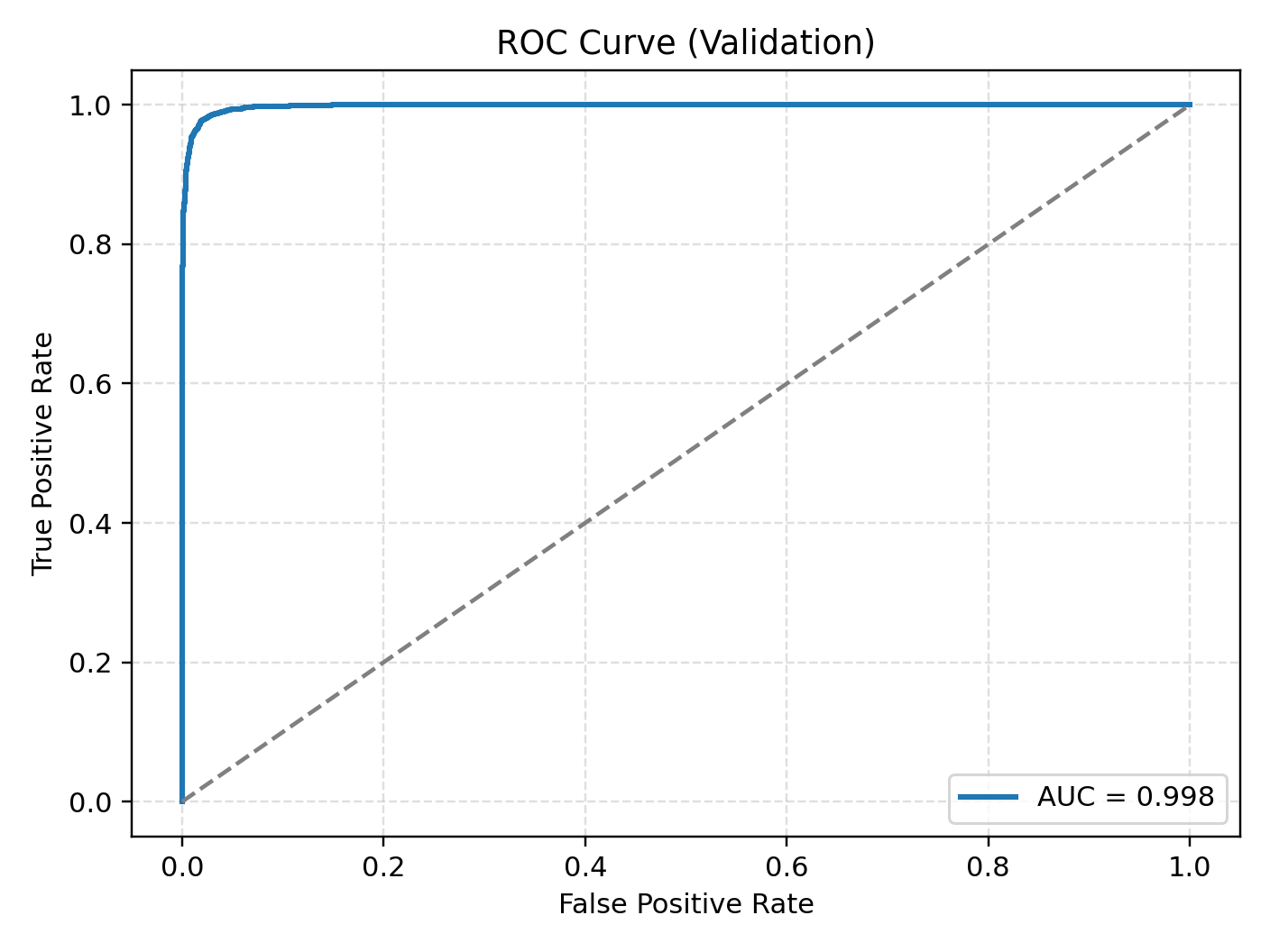} 
    \caption{
    The ROC curve summarizes how the classifier separates
    physically valid from invalid EoSs across decision thresholds. A trajectory that
    rises sharply toward the upper-left corner indicates that valid models receive
    consistently higher predicted probabilities than invalid ones. The resulting
    AUC $\approx 0.998$ shows that the surrogate captures the underlying physical
    separability of the EoS space, reflecting the distinct patterns imposed by
    stability, causal sound-speed behavior, and maximum-mass limits.
    }
    \label{fig:roc}
\end{figure}

The classification branch of the surrogate was evaluated in terms of its ability to discriminate between valid and invalid equations of state (EoSs). The Receiver Operating Characteristic (ROC) curve, shown in Figure~\ref{fig:roc}, was computed on the validation split and exhibits an Area Under the Curve (AUC) of approximately $0.997$, indicating excellent separability. The ROC curve quantifies the trade-off between the True Positive Rate (TPR) and the False Positive Rate (FPR) across a range of decision thresholds, with the upper-left corner of the plot representing the optimal classification regime.

For both the classification and regression outputs, we further evaluated the effect of the conformal prediction (CP) calibration level $\alpha$ on empirical coverage and interval tightness. Table~\ref{tab:alpha_sweep} presents these results for the standard and Mondrian CP variants. As expected, smaller values of $\alpha$ yield wider intervals and higher coverage, while larger $\alpha$ produce narrower intervals at the expense of a slight coverage reduction (Table~\ref{tab:alpha_sweep}). Across all $\alpha$, Mondrian CP attains classification accuracy comparable to or slightly higher than Standard CP. For regression, Mondrian coverage is similar to Standard, with intervals that are \emph{narrower only at} $\alpha=0.05$ and otherwise slightly wider for $\alpha\!\ge\!0.075$, likely reflecting stratification-induced sample sizes within calibration bins. Overall, both methods track the nominal $1-\alpha$ target closely (Figures~\ref{fig:cov_alpha}, \ref{fig:width_alpha}). 

\subsection{Coverage and efficiency vs.\ $\alpha$}

\begin{figure}[H]
    \centering
    \includegraphics[width=0.8\linewidth]{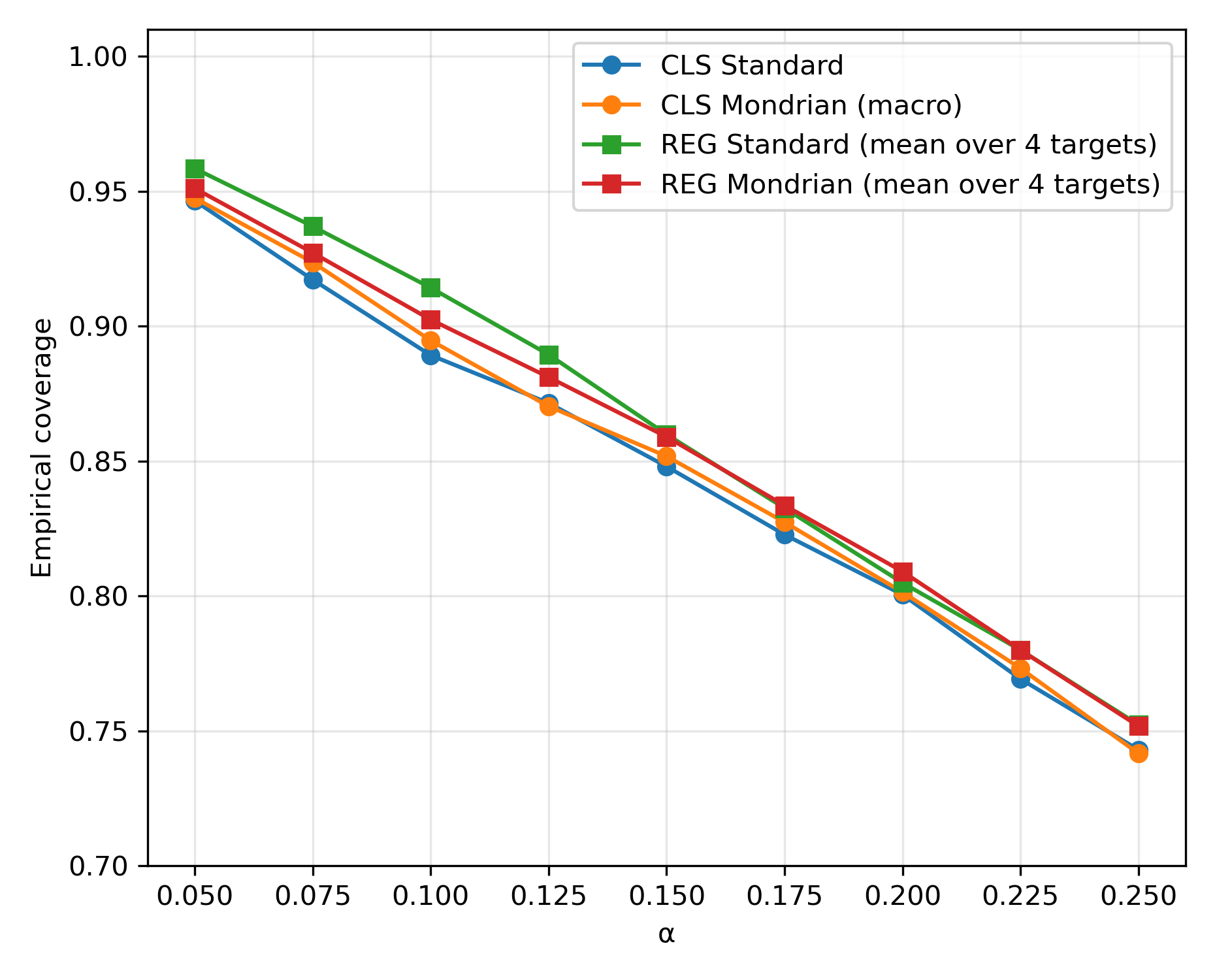}
    \caption{
    Empirical coverage for classification and regression under Standard and Mondrian
    CP as a function of $\alpha$. All curves track the nominal $1-\alpha$ within
    sampling noise, indicating that the calibrated quantiles correctly characterize
    the residual distribution. Mondrian's stratified calibration attains coverage
    statistically indistinguishable from Standard across the full range.
    }
    \label{fig:cov_alpha}
\end{figure}

\begin{figure}[H]
    \centering
    \includegraphics[width=0.8\linewidth]{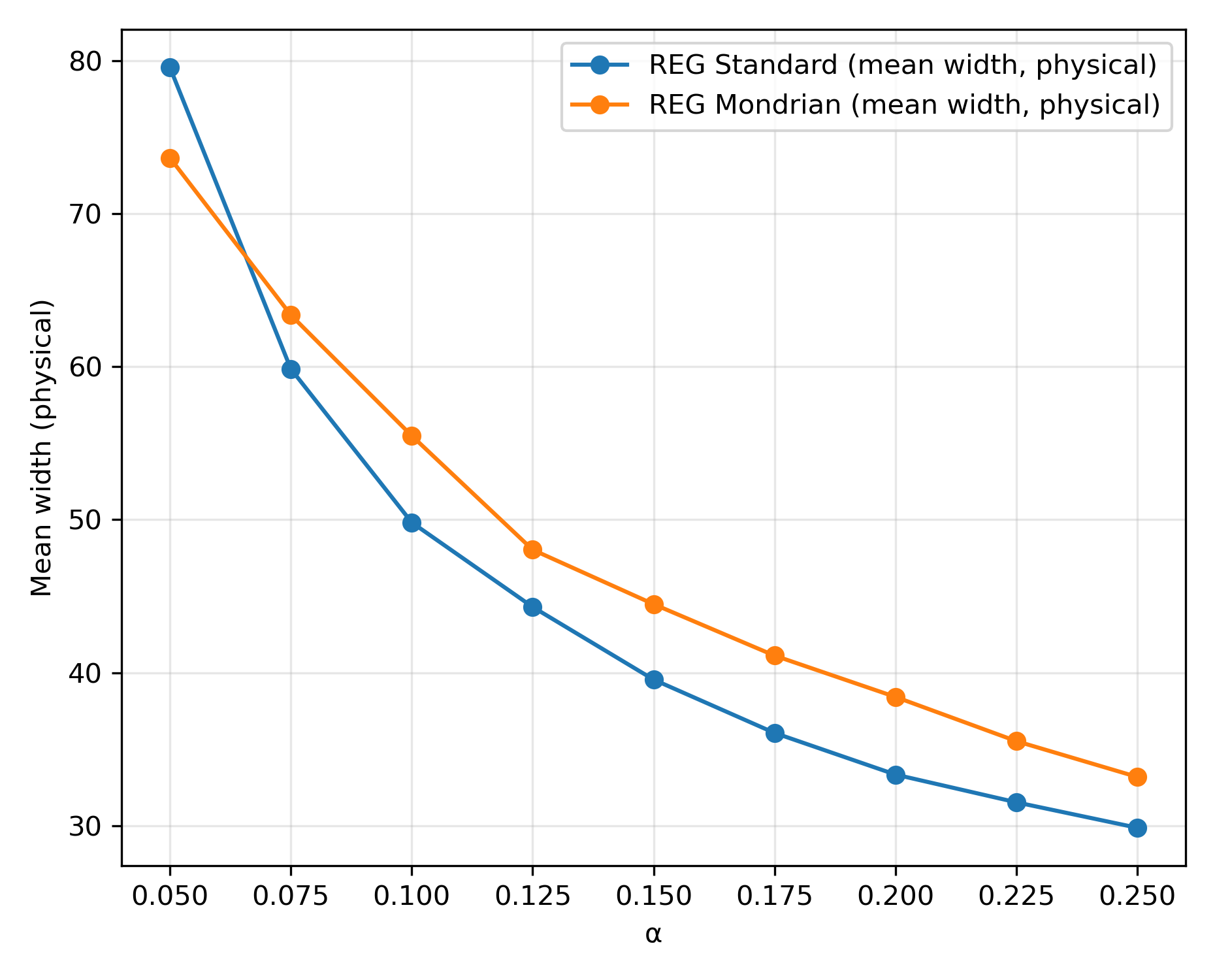}
    \caption{
    Mean prediction–interval width for the four regression targets as a function of
    $\alpha$. Interval widths decrease monotonically as the acceptance threshold
    relaxes. Mondrian yields narrower intervals when class–conditioned residuals
    are less variable, reflecting improved efficiency without loss of empirical
    coverage. Width reductions translate directly into tighter astrophysical
    constraints (e.g., km-level gains in $R_{1.4}$ or tens of units in
    $\Lambda_{1.4}$).
    }
    \label{fig:width_alpha}
\end{figure}

Figures~\ref{fig:cov_alpha}--\ref{fig:width_alpha} summarize how the surrogate’s
uncertainty intervals respond to changes in the target significance level $\alpha$.
In Fig.~\ref{fig:cov_alpha}, the empirical coverage follows the nominal
$1-\alpha$ across $\alpha\in[0.05,0.25]$, indicating that the calibrated quantiles
correctly reflect the distribution of residuals for both Standard and Mondrian CP.
The small fluctuations around the ideal line arise from finite-sample variability,
not systematic miscalibration.

Figure~\ref{fig:width_alpha} shows the corresponding interval widths in physical
units. As $\alpha$ increases, intervals shrink monotonically, reflecting the usual
coverage–efficiency trade-off. Mondrian CP yields tighter bounds when the residual
distribution differs between validity classes, because stratification reduces
within-group variance. Operationally, the reduction in width is physically meaningful:
for example, narrowing the $R_{1.4}$ interval by a few kilometers or the
$\Lambda_{1.4}$ interval by tens of units directly sharpens the discrimination
between competing EoSs, especially where observational uncertainties are comparable
in magnitude.

\begin{table*}[ht]
\centering
\small
\caption{
Sweep over significance levels $\alpha$ comparing Standard and Mondrian conformal
prediction. Metrics: (i) classification accuracy, (ii) regression coverage, and
(iii) prediction–interval width in physical units (mean across the four targets).
}
\label{tab:alpha_sweep}
\begin{tabular}{ccccccc}
\hline
$\alpha$ &
\multicolumn{2}{c}{\textbf{Cls. Accuracy}} &
\multicolumn{2}{c}{\textbf{Reg. Coverage}} &
\multicolumn{2}{c}{\textbf{Reg. Width}} \\
 & Std & Mond & Std & Mond & Std & Mond \\
\hline
0.050 & 0.949 & 0.944 & 0.957 & 0.953 & 79.0 & 73.6 \\
0.075 & 0.915 & 0.922 & 0.933 & 0.928 & 61.0 & 63.8 \\
0.100 & 0.891 & 0.896 & 0.912 & 0.902 & 50.7 & 55.1 \\
0.125 & 0.868 & 0.870 & 0.885 & 0.879 & 43.7 & 50.5 \\
0.150 & 0.845 & 0.846 & 0.861 & 0.857 & 39.9 & 46.9 \\
0.175 & 0.821 & 0.823 & 0.834 & 0.833 & 36.5 & 42.4 \\
0.200 & 0.798 & 0.799 & 0.810 & 0.809 & 34.2 & 39.0 \\
0.225 & 0.774 & 0.774 & 0.783 & 0.781 & 32.3 & 36.3 \\
0.250 & 0.746 & 0.747 & 0.758 & 0.756 & 30.7 & 34.4 \\
\hline
\end{tabular}
\end{table*}

\subsection{Absolute and Relative Errors}
\begin{figure}[H]
  \centering
  \includegraphics[width=0.8\linewidth]{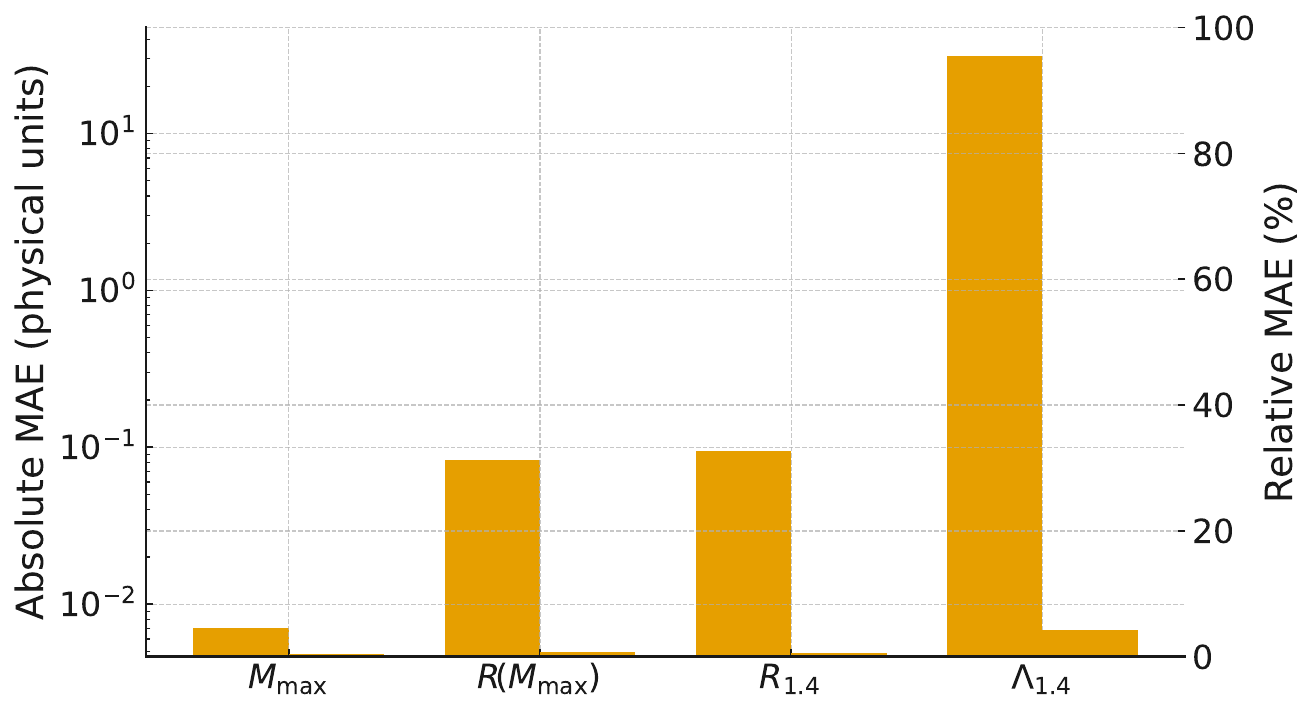}
  \caption{\textbf{Validation MAE (absolute and relative).}
  Absolute MAE (left axis, log scale) and relative MAE (right axis) for the four scalar targets. 
  Masses and radii show sub-percent errors; $\Lambda_{1.4}$ displays a few-percent MAE, consistent with its $R^5$ sensitivity.
  }
  \label{fig:val_mae}
\end{figure}

\begin{figure}[H]
  \centering
  \includegraphics[width=0.8\linewidth]{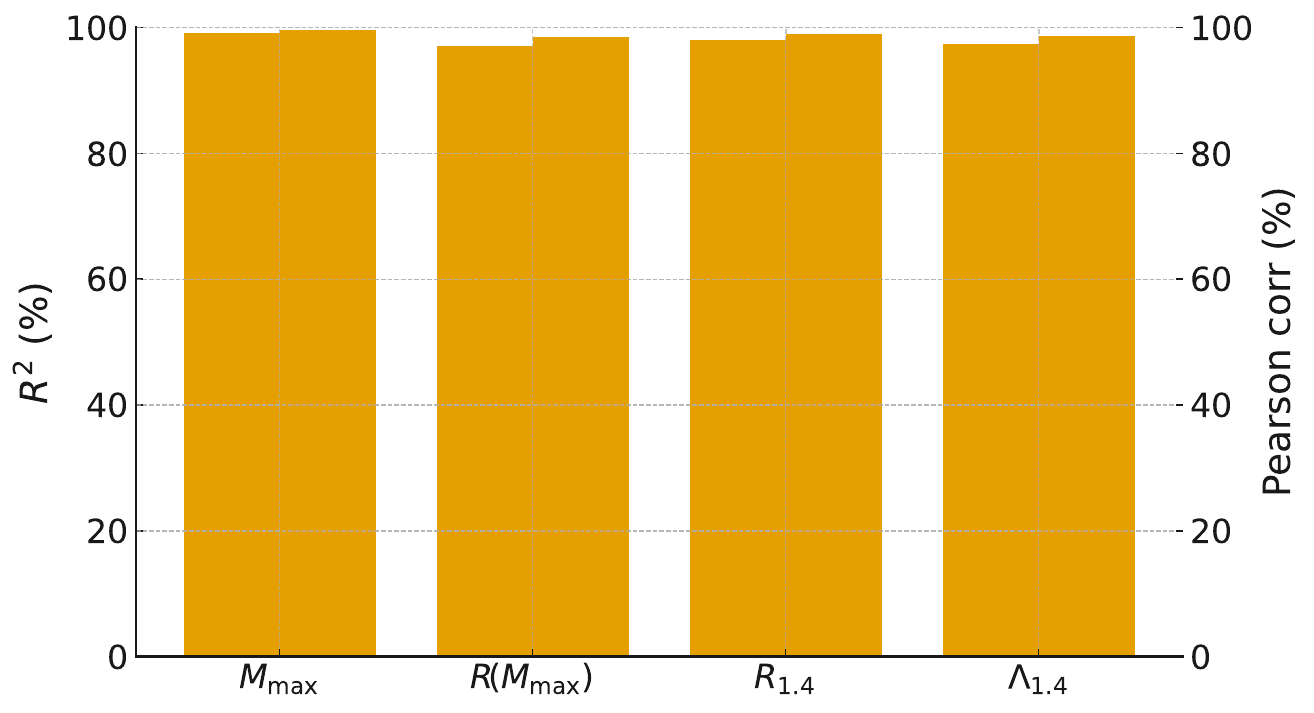}
  \caption{\textbf{Validation $R^2$ and Pearson correlation.}
  The surrogate explains nearly all variance for $M_{\max}$ and radii ($>\!97\%$) and maintains strong predictive power for $\Lambda_{1.4}$ ($\sim\!97\%$), despite its enhanced physical sensitivity.
  }
  \label{fig:val_r2corr}
\end{figure}

We assess point-prediction accuracy using the Mean Absolute Error (MAE) and the coefficient of determination ($R^{2}$), shown in Figures~\ref{fig:val_mae} and \ref{fig:val_r2corr}. These metrics jointly quantify the typical physical deviation from full TOV solutions and the fraction of variance explained for each observable. As seen in Figure~\ref{fig:val_mae}, the surrogate achieves sub-percent accuracy for the maximum mass and stellar radii ($M_{\max}\!\sim\!0.02\,M_\odot$ and $R(M_{\max}),R_{1.4}\!\sim\!0.1\,$km), well below current NICER uncertainties and therefore negligible for astrophysical inference. The tidal deformability $\Lambda_{1.4}$ exhibits larger dispersion (MAE $\sim$31, $\sim$4\%), consistent with the strong sensitivity $\Lambda\propto(R/M)^5$, which amplifies small radius variations, especially in the high-density regime where EoS variability is intrinsically larger.

Figure~\ref{fig:val_r2corr} provides the corresponding scale-independent view: $R^{2}\approx 99\%$ for $M_{\max}$ and $>97\%$ for both radii, indicating that the surrogate captures nearly the entire variance of these low-to-intermediate density–dominated quantities. Even $\Lambda_{1.4}$, despite its nonlinear dependence on the radius, reaches $R^{2}\approx 97\%$ with high correlation, fully consistent with the few-percent MAE. Physically, these results show that masses and radii—set primarily by the bulk EoS—are reconstructed with excellent precision, while the remaining uncertainty in tidal deformability remains small compared to current GW intervals, which span several hundred units. Overall, the surrogate provides astrophysically robust fidelity across all scalar targets, with modeling errors well below the precision of present multimessenger observations.

\subsection{External Validation with an Independent Dataset}
\label{subsec:external_validation_distinct}

We performed an external validation using a fully independent dataset (\emph{Ext-Set}) generated only after the surrogate---including architecture, weights, scaler, and conformal calibration—was frozen. This ensures a strict separation between development and evaluation, preventing information leakage and providing a genuine test of out-of-distribution (OOD) generalization.

The \emph{Ext-Set} was constructed with: (i) a strict no-overlap policy using hash fingerprints of each EoS; (ii) shifted sampling ranges for polytropic indices and transition densities to induce mild distributional shift; (iii) independent preprocessing (filters and sanity checks) performed outside the development pipeline; and (iv) blind TOV labels computed prior to any surrogate evaluation. All predictions used the fixed threshold of $0.99$ on the validity probability, with no re-tuning or recalibration.

Table~\ref{tab:ext_validation} summarizes the comparison between surrogate predictions and TOV outcomes. The classifier maintains strong agreement under distributional shift, with misclassifications confined to borderline cases. This indicates that the learned decision boundary—driven by maximum-mass, stability, and tidal/radius constraints—remains robust outside the training distribution.

\begin{figure}[H]
    \centering
    \includegraphics[width=0.8\linewidth]{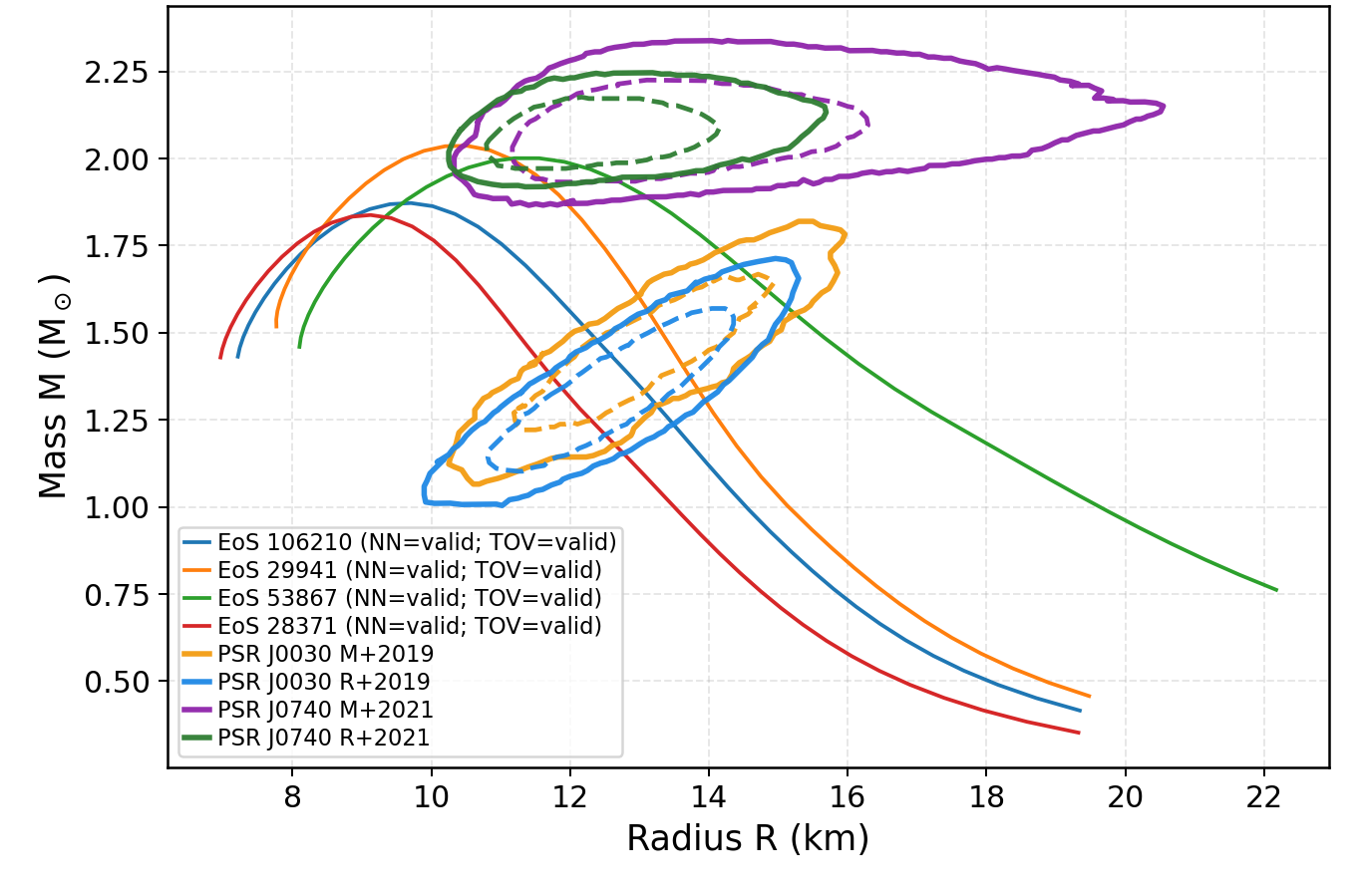}
    \caption{$M$--$R$ curves for four externally validated EoSs with NN=valid and TOV=valid. 
    Solid curves correspond to full TOV integrations, while the NICER credible regions for 
    PSR~J0030 and PSR~J0740 are overplotted as observational constraints 
    \citealt{Miller2019_ApJL_J0030,Riley2019_ApJL_J0030,
    Miller2021_ApJL_J0740,Riley2021_ApJL_J0740}. 
    All tracks intersect the allowed observational domains and reach 
    $M_{\max}\!>\!2.0\,M_\odot$, supporting their physical plausibility.}
    \label{fig:top4_mr}
\end{figure}

To illustrate physical consistency, Figure~\ref{fig:top4_mr} shows the $M$--$R$ curves for four externally validated EoSs for which both surrogate and TOV label agree (\emph{NN=valid; TOV=valid}). Their tracks intersect the NICER credible regions for PSR~J0030$+$0451 and PSR~J0740$+$6620 and reach $M_{\max}\!>\!2.0\,M_\odot$, confirming that the surrogate's ``valid'' predictions correspond to astrophysically plausible solutions aligned with independent multimessenger constraints.

Overall, the surrogate generalizes reliably: when it predicts validity, the underlying TOV integrations produce physically consistent $M$--$R$ curves that satisfy observational bounds. This demonstrates that the model is suitable for rapid screening, large-scale sweeps, and downstream inference where fast, physically grounded evaluation is required.

\begin{table}[t]
\centering
\caption{External validation results comparing surrogate predictions (RN) against TOV outcomes.}
\label{tab:ext_validation}
\begin{tabular}{lcc}
\hline
                & TOV: Invalid & TOV: Valid \\
\hline
RN: Invalid     & 39           & 5          \\
RN: Valid       & 6            & 50         \\
\hline
\end{tabular}

\vspace{2pt}
\small Threshold fixed at $p_{\mathrm{valid}}>0.99$. Most cases agree with TOV, with only a few borderline discrepancies.
\end{table}

\subsection{Benchmarking Against Related Work}
\label{sec:benchmarks}

To situate the surrogate within the broader landscape of neutron-star equation-of-state (EoS) inference, it is useful to contrast its performance with representative Bayesian, nonparametric, and neural-surrogate approaches. Because these methodologies differ fundamentally in modelling assumptions, data structures, and interpretations of uncertainty, the comparison is contextual rather than metric-by-metric. The focus is on the scale of predictive errors and on the nature of the uncertainty guarantees.

Table~\ref{tab:comparison_bench_expanded} summarizes the present model’s accuracy under class-conditioned conformal prediction. The surrogate achieves mean absolute errors of $\sim0.02\,M_\odot$ for the maximum mass, $\sim 0.1$ km for $R_{1.4}$, and $\sim 31$ for $\Lambda_{1.4}$, corresponding to sub-percent errors in masses and radii and few-percent in tidal deformability. These values are competitive with, and in several cases tighter than, the widths typically reported in Bayesian posterior intervals.

\begin{table}[t]
\centering
\caption{Performance of the proposed surrogate with class-conditioned conformal prediction.}
\label{tab:comparison_bench_expanded}

\resizebox{\linewidth}{!}{%
\begin{tabular}{lcccc}
\hline
\hline
\textbf{Work} &
\textbf{Domain / Inputs} &
\textbf{Outputs} &
\textbf{Uncertainty / Error} &
\textbf{Notes} \\
\hline
This work (MAE) & Piecewise--polytropic NN + CP & $M_{\max}$      
& $\approx 0.02\,M_\odot$ ($\sim$0.4\%) & Certified coverage (CP) \\

This work (MAE) & same                          
& $R_{1.4}$       
& $\approx 0.1$ km ($\sim$0.7\%)        & --- \\

This work (MAE) & same                          
& $\Lambda_{1.4}$ 
& $\approx 31$ ($\sim$4.3\%)            & --- \\
\hline
\end{tabular}%
}

\vspace{2pt}
\footnotesize
Mondrian conformal prediction yields distribution-free coverage conditioned on class structure.
\end{table}

Bayesian multimessenger analyses such as \citet{Han2021} report posterior credible intervals for radii and tidal deformabilities under explicit prior and likelihood models. Nonparametric or hybrid approaches, e.g. \citet{Zhou2023}, provide flexible posterior bounds for quantities such as $M_{\max}$. Fast neural surrogates in relativistic astrophysics and modified-gravity settings \citep{Liodis2024,Danarianto2025} emphasize computational throughput but generally do not supply formally calibrated uncertainty. These works together delineate the three principal paradigms in EoS inference: full Bayesian posteriors, nonparametric or hybrid distributions, and deterministic neural accelerators.

Although the uncertainty notions differ substantially—Bayesian credible intervals quantify posterior probability mass, while conformal intervals guarantee empirical coverage $1-\alpha$ independently of model correctness—placing them side by side indicates scale compatibility. The surrogate’s residuals for $M_{\max}$ and radii fall below the widths of several-percent credible intervals commonly quoted in Bayesian analyses, whereas $\Lambda_{1.4}$ errors remain within a few percent, well below current multimessenger uncertainty budgets. Conformal prediction thus provides a complementary mode of uncertainty quantification: a frequentist, distribution-free guarantee that remains valid under moderate distributional shift and does not rely on explicit prior assumptions.

Absolute numbers across studies inevitably depend on choices such as EoS parameterization, multimessenger data combinations, and reporting conventions (e.g., 90\% versus 95\% credibility). To minimize such effects, this work reports physical errors in parallel with relative percentages and explicitly specifies the conformal level $1-\alpha$. Under these conditions, the surrogate attains high accuracy across all observables while providing uncertainty sets that are formally calibrated in finite samples.

In summary, the proposed surrogate achieves predictive accuracy competitive with state-of-the-art Bayesian and nonparametric EoS inference methods, while introducing a capability largely absent from fast neural surrogates: distribution-free, class-conditioned uncertainty guarantees. This combination of high throughput, low bias, and certified coverage positions the approach as a statistically robust alternative to full TOV pipelines in large-scale or iterative inference settings.

\section{Discussion and Conclusions}\label{sec:disc_conc}

This work introduces the first surrogate for neutron-star equation-of-state (EoS) inference
explicitly calibrated with both Standard and Mondrian conformal prediction (CP), providing
distribution-free and finite-sample uncertainty guarantees. By combining multitask learning
with CP, the surrogate delivers reliable uncertainty control without assuming Bayesian priors
or parametric likelihoods, offering a complementary alternative to traditional inference
pipelines.

The model achieves high predictive accuracy: ROC analysis shows near-perfect discrimination
between physically valid and invalid EoSs, while empirical CP coverage closely follows the
nominal $1-\alpha$ across all tested significance levels. Because inference requires only a
single neural-network evaluation and a CP quantile lookup, the approach remains computation\-ally
lightweight and scalable to large parameter sweeps.

A central impact of this framework lies in its ability to interface cleanly with modern
observational constraints. The surrogate reproduces $M_{\max}$, radii, and tidal deformabilities
with errors well below current NICER credible-region widths and far smaller than the
$\Lambda(M)$ uncertainties derived from LIGO/Virgo binary-merger events. Consequently, the
model's residuals do not compete with---and therefore do not bias---astrophysical constraints.
Instead, the calibrated CP intervals provide statistically controlled envelopes that can be
directly juxtaposed with NICER posteriors and GW-based bounds on $\Lambda_{1.4}$, enabling
rapid and reliable mapping between microphysical assumptions and observationally allowed
regions.

Limitations remain. Because the surrogate is trained on three-segment piecewise--polytropic
generators in the spirit of \citet{Read2009}, it cannot fully capture sharp phase transitions
or exotic degrees of freedom. CP guarantees hold marginally under exchangeability and do not
enforce physical structure (e.g., monotonicity or causality) at the surrogate level. Moreover,
the present targets are limited to scalar observables rather than full functional curves.

Future developments will therefore focus on \emph{functional surrogates} capable of learning
the entire $R(M)$ or $\epsilon(P)$ relations using sequence models such as Transformers or
temporal convolutional networks, paired with physics-informed decoders and CP calibration in
function space. Such models would allow direct comparison against full NICER radius
posteriors and LIGO/Virgo $\Lambda(M)$ constraints, moving beyond pointwise quantities
and enabling holistic consistency checks across the mass range.

In summary, this study establishes a reproducible and extensible framework in which fast
neural surrogates are endowed with formally calibrated uncertainties. By introducing conformal
prediction into neutron-star EoS modeling for the first time, we bridge efficient machine
learning with rigorous uncertainty quantification, providing a scientifically reliable tool
for current and forthcoming multimessenger constraints.

\section*{Acknowledgments}
The authors acknowledge support from the Instituto Tecnológico de Aeronáutica (ITA) for providing the research environment and computational resources that enabled this work. M.M.S.M. thanks the Coordenação de Aperfeiçoamento de Pessoal de Nível Superior (CAPES), Brazil, for financial support. This work has been done as a part of the Project INCT-F\'isica Nuclear e Aplica\c{c}\~oes, under No. 408419/2024-5. It is also supported by Conselho Nacional de Desenvolvimento Cient\'ifico e Tecnol\'ogico (CNPq) under Grants No. 301779/2025-2 (M.D.), No. 305327/2023-2 (C.H.L.), No. 401565/2023-8~(Universal - C.H.L., M.D.), No. 409736/2025-2~(Universal - C.H.L., M.D.), No. 444797/2024-6 (M.D.), and Funda\c{c}\~ao de Amparo \`a Pesquisa do Estado de S\~ao Paulo (FAPESP) under Thematic Project No. 2024/17816-8 (C.H.L., M.D.).

\bibliographystyle{aasjournal}
\bibliography{refs}

\end{document}